\documentclass[conference]{IEEEtran}
\IEEEoverridecommandlockouts

\usepackage{graphicx}
\usepackage{caption}
\usepackage{subcaption}
\usepackage{url}
\usepackage{amsfonts}
\usepackage{amsmath}
\usepackage{amssymb}

\usepackage[colorlinks=true,citecolor=blue,linkcolor=blue,urlcolor=blue]{hyperref}

\usepackage[sorting=none]{biblatex}
\makeatletter
\newcommand{\currentfontsize}{The current font size is: \f@size pt}
\newcommand{\currenttextwidth}{The current text width is: \the\textwidth}
\newcommand{\currentcolumnwidth}{The current column width is: \the\columnwidth}
\makeatother

\newcommand\figsize{0.5}
\newcommand\fullfigsize{0.9}

\begin{document}

\title{
    Extending UNIQuE: Quantum Simulation Speedup for the HHL Algorithm
}
\author{
    \IEEEauthorblockN{Reece Robertson}
    \IEEEauthorblockA{
        \textit{Department of Computer Science \& Electrical Engineering} \\
        \textit{Department of Physics} \\
        \textit{Quantum Science Institute} \\
        \textit{University of Maryland, Baltimore County} \\
        Baltimore, USA \\
        reecerobertson@umbc.edu \\
        0000-0003-1064-0012
    }
    \and
    \IEEEauthorblockN{Ameya S. Bhave}
    \IEEEauthorblockA{
        \textit{Department of Electrical and Computer Engineering} \\
        \textit{The University of Texas at Dallas} \\
        Richardson, TX, USA \\
        asb240006@utdallas.edu \\
        0009-0006-7355-4807
    }
}

\date{\today}

\maketitle

\begin{abstract}
    In an extension of the Unconventional Noiseless Intermediate Quantum Emulator, this work introduces a classical emulation of the quantum Harrow-Hassidim-Lloyd algorithm for sampling from the solution space of linear systems.
    The emulated HHL algorithm scales exponentially with the number of qubits required to represent the linear system, which is an advantage over the state vector simulation of the HHL algorithm, which scales exponentially as a function of both the size of the linear system and the magnitude of its largest (scaled) eigenvalue.
    We benchmark our emulator by comparing it with the Intel Quantum Simulator and demonstrate a runtime advantage for small linear systems.
\end{abstract}

\begin{IEEEkeywords}
    Quantum Computing, Quantum Simulation, Linear Systems
\end{IEEEkeywords}

\section{Introduction}




Quantum computers may become an important component of tomorrow's supercomputing ecosystem.
Several quantum algorithms have been discovered that solve certain problems with provable speedups over the corresponding fastest known classical methods for the same problems \cite{Dalzell_2025}.
The speedups are often predicated on the assumption that one has access to a quantum system over which one has perfect control (i.e., a noiseless quantum computer).
One such algorithm is the Harrow, Hassidim, and Lloyd (HHL) algorithm, which samples from the solution to a system of linear equations and (under favorable conditions) gains an exponential speedup over the comparable classical algorithm \cite{harrow_quantum_2009}.
This algorithm will be reviewed in more detail in Section \ref{hhl-algorithm}.

Ironically, the current state-of-the-art for small-scale \textit{error-free} quantum computation is to avoid real quantum computers altogether---we have not yet attained perfect control over any real quantum system \cite{Knill2005, Resch_Benchmarking_2021}.
Moreover, the noise observed in current quantum devices often severely impacts the performance of algorithms executed on these devices (see, for example, \cite{Robertson_Simons_2025, Robertson_Grovers_2025, Robertson_LightRail_2025, Robertson_Gibbs_2026}).
On real quantum computers, the theory of quantum error correction, and related techniques, can be used to suppress errors \cite{lidar2013quantum}.
Error suppression has been successfully demonstrated on real hardware \cite{Acharya2025, Robertson_BreakEven_2026, Robertson_Steane_2025}, yet truly noiseless quantum computation has not yet been achieved.

To circumvent the issue of noise, researchers have developed methods to simulate quantum algorithms (including the HHL algorithm) on classical computers using matrix operations.
Several types of simulation techniques exist, including state vector simulators, density matrix simulators, and tensor network simulators.
State vector simulators model a quantum state using a vector, and evolve the state through a quantum algorithm using matrix-vector calculations \cite{Barenco_Elementary_1995, nielsen_quantum_2010}.
Density matrix simulators use a matrix to represent a quantum state, which requires more resources than state vector simulation but allows for more expressive modeling of quantum noise processes \cite{berendsen1993quantum}.
Tensor networks can efficiently simulate large quantum algorithms using a network of tensors.
These simulations can be exact or approximate, with increasingly large speedups attained at the cost of decreasingly exact approximations \cite{Orus2019}.

In this work, we improve upon the HHL simulation using optimized numerical methods.
This represents an extension of the Unconventional Noiseless Intermediate Quantum Emulator (UNIQuE), introduced in \cite{Robertson_UNIQuE_2025}.
Both the HHL simulation and emulation scale exponentially with the number of qubits required to represent the linear system, yet the emulation avoids an exponential dependence incurred by the simulation on the number of qubits devoted to the accuracy of the approximation.
We benchmark our routine against the Intel Quantum Simulator, a leading state vector simulator, to show an advantage.
We choose to compare our results to a state vector simulator, as these simulators perform relatively efficient exact simulation of noiseless quantum algorithms.
We avoid density matrix simulation as we are concerned with noiseless (not noisy) quantum simulation.
We likewise avoid tensor network simulation as we are concerned with exact (not approximate) quantum simulation.
The details of the state vector simulation are given in Section \ref{simulation}, which is contrasted to our novel approach in Section \ref{emulation}.

A subtlety to note is that in this paper, we are not concerned with solving a linear system of equations, per se.
Instead, we are concerned with replicating the behavior of the HHL quantum algorithm on a classical computer.
Quantum computers solve problems by creating and sampling from specific distributions from which meaningful information can be extracted \cite{nielsen_quantum_2010, Mermin_2007}.
As such, we are here concerned with preparing and sampling from this same distribution on a classical computer as efficiently as possible.
To this end, we introduce an emulation of the HHL algorithm, and we benchmark it against a simulation of the same algorithm.
We show that both methods ultimately produce and sample from the distribution that would be prepared by a perfect quantum computer executing the HHL algorithm, and then we show that our emulation achieves significantly better scaling than the standard simulation method.

We close the introduction with a word of terminology.
Throughout the paper, we use the term \textit{simulation} to refer to the standard (state vector) method for simulating quantum algorithms.
We use the term \textit{emulation} to describe our method of simulating the algorithm directly as efficiently as possible using an optimized algorithm.\footnote{See the appendix for a Python implementation of the emulation routine for a $2\times2$ matrix $A$.}
We note that this represents some departure from the usual usage of these words; usually, an \textit{emulator} is a more faithful model of the underlying physics of a device or process than a \textit{simulator}.
In our work, the inverse is true: our algorithm-level emulator is more efficient, but more abstract, than our simulator.
Nevertheless, we persist with the usage to remain consistent with prior usage of these terms in the field of quantum information processing \cite{Robertson_UNIQuE_2025, Haner_Emulation_2016}.
 
\section{Methods} \label{description}

We begin with a discussion of how the HHL algorithm works on a quantum computer.
Following this, we describe in detail how we implement this on a classical computer via state vector simulation, and then via our emulation method.

\subsection{The HHL Algorithm on a Quantum Computer}\label{hhl-algorithm}

\begin{figure*}[t]
    \centering
    \includegraphics[width=\fullfigsize\linewidth]{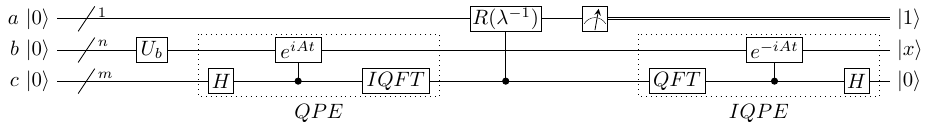}
    \caption{
        A high-level overview of the HHL algorithm.
        The algorithm uses three quantum registers: a one-qubit \textit{ancilla} register $a$, an $n$-qubit $b$ register to encode the value of $|b\rangle$ (and later $|x\rangle$), and an $m$-qubit \textit{clock} register $c$.
        The algorithm begins by encoding $|b\rangle$ into $b$ via the state-preparation unitary $U_b$.
        Next, the Quantum Phase Estimation (QPE) routine is used to apply the eigenvalues of $A$ to $b$.
        This routine uses the Inverse Quantum Fourier Transform (IQFT) as a subroutine.
        Following this, an $R(\lambda^{-1})$ rotation is used to invert the eigenvalues applied to $b$ (provided that $a$ is measured to be in the $|1\rangle$ state), and the Inverse Quantum Phase Estimation (IQPE) is used to disentangle $c$ from $b$.
        In the end, the solution $|x\rangle$, as shown in \eqref{postMeasurement}, is encoded in the $b$ register.
        For more details, see \cite{harrow_quantum_2009, morrell2023stepbystep}.
    }
    \label{overview}
\end{figure*}

The HHL algorithm begins with a system of linear equations $A|x\rangle=|b\rangle$, where $A$ is an $s$-sparse $N\times N$ invertible Hermitian matrix, $|b\rangle$ is an $N\times1$ unit vector, and $|x\rangle$ is the unknown solution vector to this equation.
We also have that $N=2^n$ for some $n\in\mathbb{N}$, which implies that $|b\rangle$ can be encoded in the state of $n$ qubits.
Let $|u_j\rangle$ be the eigenvectors of $A$ with $\lambda_j$ as the corresponding eigenvalues, that is,
\begin{equation}
    A|u_j\rangle = \lambda_j|u_j\rangle
\end{equation}
for all $j\in\{1, \ldots , N\}$.
Given the correct coefficients $\beta_j$, we can represent $|b\rangle$ in the eigenbasis of $A$ as
\begin{equation}
    |b\rangle=\sum_{j=1}^N \beta_j|u_j\rangle.
\end{equation}
Now, as stated in \cite{harrow_quantum_2009}, we have that
\begin{equation}
    \sum_{j=1}^N \beta_j \lambda_j^{-1} |u_j\rangle = A^{-1}|b\rangle = |x\rangle.
    \label{solution}
\end{equation}
Our quantum computer produces this solution by preparing the state
\begin{equation}
    \sum_{j=1}^N \beta_j |u_j\rangle \left(\sqrt{1 - \frac{C^2}{\lambda_j^2}}|0\rangle + \frac{C}{\lambda_j}|1\rangle\right),
    \label{equation}
\end{equation}
which, upon measuring a $|1\rangle$ in the second register, becomes
\begin{equation}
    K\sum_{j=1}^N \beta_j \frac{C}{\lambda_j} |u_j\rangle,
    \label{postMeasurement}
\end{equation}
where $C$ and $K$ are normalizing constants.\footnote{For details, consult \cite{harrow_quantum_2009}.}
Observe that \eqref{postMeasurement} is identical to \eqref{solution} up to the presence of the constants $CK$, and hence \eqref{postMeasurement} represents a (scaled) solution to our problem.

The HHL algorithm prepares \eqref{solution} on a quantum computer with an expected runtime that scales as $O(\log(N)s^2\kappa^2/\epsilon)$, where
\begin{equation}
    \kappa = \frac{\lambda_{\max}}{\lambda_{\min}}
\end{equation}
is the condition number of the Hermitian matrix $A$ \cite{harrow_quantum_2009}, and $\epsilon$ sets the error tolerance.
This provides exponential advantage over the conjugate gradient method, which has a runtime that scales as $O(N\sqrt{\kappa})$ \cite{shewchuk1994introduction}.
More saliently for our purposes, the complexity of preparing the distribution from which the ideal HHL algorithm samples scales as $\text{poly}(\log(N),\kappa)$.
When discussing the complexity of simulating and emulating the HHL algorithm below, we will report the complexity of preparing this distribution.\footnote{We choose to consider the complexity of preparing the distribution of the HHL algorithm because, when performing the algorithm on a classical computer, once the state vector has been generated, it can be efficiently sampled without rerunning the entire routine.}

With a mind toward brevity, we do not describe the details of the algorithm here; those interested should consult the original paper by Harrow, Hassidim, and Lloyd \cite{harrow_quantum_2009}, or the very readable introduction by Morrell et al. \cite{morrell2023stepbystep}.
Suffice it to say that the crux of the algorithm lies in applying the inverse eigenvalues of $A$ to the state $|b\rangle$ \cite{harrow_quantum_2009}.
This is done using the quantum phase estimation (QPE) subroutine, which involves converting the Hermitian matrix $A$ into a unitary matrix $\exp(iAt)$ and repeatedly applying it to evolve the quantum state $|b\rangle$.
A high-level picture of the algorithm, presented in the language of quantum circuit diagrams, is given in Fig. \ref{overview}.

One last point to note is that the algorithm makes use of three quantum registers, labeled $a$, $c$, and $b$ from top to bottom in Fig. \ref{overview}.
Each quantum register contains some number of qubits, and each additional qubit doubles the representative capacity of the register.
Register $a$ always consists of a single \textit{ancilla} qubit.
Register $b$ is initialized to the state $|b\rangle$ (hence its name) and is evolved into the state $|x\rangle$ over the course of the algorithm.
Thus, this register needs to contain $n=\log_2(N)$ qubits, which is enough to represent the $N\times1$ vector $|b\rangle$.
In our experiments below, we fix the size of the $b$ register at 1 qubit, which allows for $|b\rangle$ vectors of size $2\times1$.
Finally, the $c$ register represents the \textit{clock} qubits.
This register requires enough qubits to accurately approximate the ratio between the eigenvalues of $A$; that is, if $\lambda_{\min}:\lambda_{\max}$ reduces to $r_{\min}:r_{\max}$, then the $c$ register should contain
\begin{equation}
    m = \lceil \log_2(r_{\max}+1) \rceil
\end{equation}
qubits to represent the ratio exactly.
If $m$ is less than this value, then some approximation error will be incurred in the algorithm.
In general, increasing the number of counting qubits increases the accuracy of the HHL algorithm until the ratio between the eigenvalues can be expressed exactly \cite{morrell2023stepbystep}.

\subsection{Simulating the HHL Algorithm}\label{simulation}

To simulate the HHL algorithm, we reproduce every operation required by the algorithm on a classical computer.
Recall that a register of $x$ qubits is representable by a state vector of size $2^x$, and a quantum gate operation applied to this register is representable by a $2^x\times2^x$ unitary matrix.

The HHL algorithm involves three registers: the one-qubit $a$ register, the $ n$-qubit $b$ register, and the $m$-qubit $c$ register.
To simulate the HHL routine, a series of matrices representing each quantum is applied to the state vector via matrix multiplication.
If a dense full-register matrix-vector implementation is used, then the matrices are of size $2^{m+n+1}\times2^{m+n+1}$ and the runtime of the routine scales as $O(N^22^{2m})$.
The precise runtime will depend upon the implementation details; a more general gate-based state-vector simulation cost would scale as $O(gN2^m)$, where $g$ is the number of elementary gates.
Regardless, the routine is $\text{poly}(N,2^m)$.

We implemented a state vector simulation of the algorithm using the Intel Quantum Simulator \cite{Guerreschi_2020} (formerly qHiPSTER \cite{smelyanskiy2016qhipsterquantumhighperformance}), which is a competitive entity in the field of local quantum state vector simulation.

\subsection{Emulating the HHL Algorithm}\label{emulation}

When it comes to emulating the HHL algorithm, we do not concern ourselves with exactly replicating every step of the algorithm as it would be performed on a quantum computer.
Rather, we are concerned with producing and sampling from the same distribution that a perfect quantum computer would produce and sample if it ran the algorithm.
For the HHL algorithm, this means that we just need to produce equation \eqref{equation} as efficiently as possible.
This focus on the measured result allows us to skip over many of the details of the algorithm, such as the quantum phase estimation, that the simulator must replicate.

The emulation proceeds by first computing the eigenvalues $\lambda_j$ and eigenvectors $|u_j\rangle$ of the matrix $A$, where $j\in[1,2]$.
The coefficients $\beta_j$ to represent vector $|b\rangle$ in the eigenbasis of $A$ are then computed.
Once we have that information, we compute \eqref{equation} directly.
Then we normalize the state and measure the ancilla qubit to get equation \eqref{postMeasurement}, and measure the $b$ qubit to get the final output.

Note that equation \eqref{equation}, and hence the above description, does not include the $c$ register at all.
While the $c$ register is a critical component to the quantum algorithm (and the simulation), it is always reset to $|0\rangle_m$ at the end of the (noiseless) algorithm.
Thus, since we aim to produce the final output of the algorithm, we can entirely omit this register from our emulation.

The dominant complexity cost of this routine is diagonalizing $A$, which approaches $O(N^3)$ in the worst case of a dense Hermitian $A$.
In practice, this operation scales as $\text{poly}(N)$ \cite{shewchuk1994introduction}.
Significantly, however, this emulating the distribution of the HHL algorithm has no exponential dependence upon $m$, which represents a significant improvement over the simulation method.

\begin{figure*}[t]
    \centering
    \begin{subfigure}{\figsize\textwidth}
        \centering
        \includegraphics[width=1\linewidth]{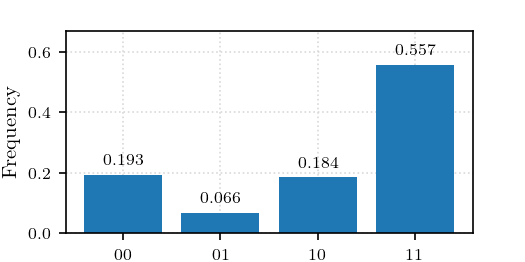}
        \caption{Simulation}
        \label{sim1}
    \end{subfigure}%
    \begin{subfigure}{\figsize\textwidth}
        \centering
        \includegraphics[width=1\linewidth]{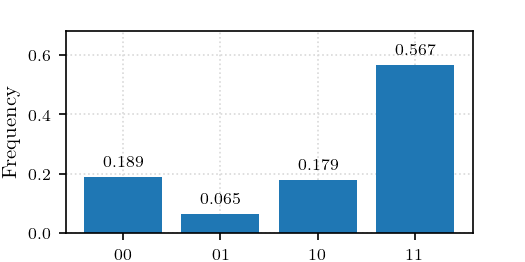}
        \caption{Emulation}
        \label{em1}
    \end{subfigure}
    \caption{Illustrative results of simulating and emulating the HHL algorithm on \eqref{ex1}.}
    \label{experiment1}
\end{figure*}

\section{Results}

We tested our emulator against our simulator on two different problems.
The first was
\begin{equation}
    \begin{pmatrix}
        1 & -1/3 \\
        -1/3 & 1
    \end{pmatrix}\begin{pmatrix}
        x_1 \\ x_2
    \end{pmatrix}=\begin{pmatrix}
        0 \\ 1
    \end{pmatrix},
    \label{ex1}
\end{equation}
chosen because it is the example used in \cite{morrell2023stepbystep}, which allowed us to verify the accuracy of our methods by reproducing the results from that paper.
In this case, the eigenvalues of $A$ are $\lambda_1=2/3$ and $\lambda_2=4/3$.
As such, the ratio between the eigenvalues is $1:2$, so $m=2$.
Hence, we performed a simulation involving 4 total qubits across all registers.

The solution to this equation is $x_1=3/8$ and $x_2=9/8$, which has a squared ratio of $1:9$.
Thus, we expect that the ratio between the probability of measuring the states where the ancilla qubit is $|1\rangle$, that is, the states $|11\rangle$ and $|01\rangle$, will also be $1:9$.\footnote{In this paper, we omit the $c$ register in all state vector representations, as it is reset to $|0\rangle_m$ at the end of the algorithm. Moreover, we use little-endian notation, hence state vectors are read as $|ba\rangle$.}
Therefore, our expected result is
\begin{equation}
    \begin{pmatrix}
        0.1 \\ 0.9
    \end{pmatrix}.
    \label{ex1sol}
\end{equation}
Moreover, we expect the probability of measuring the states where the ancilla qubit is $|0\rangle$, that is, $|10\rangle$ and $|00\rangle$, to be comparatively low.

Our expectations are verified in Fig. \ref{experiment1}, which shows a histogram derived from running 2048 shots of this problem on the simulator (Fig. \ref{sim1}) and on the emulator (Fig. \ref{em1}).
We see that both histograms are close to each other, indicating that both the simulator and emulator sample from the correct distribution at the conclusion of the algorithm.\footnote{Moreover, both histograms are very similar to Fig. 4 of \cite{morrell2023stepbystep}, which further establishes that both our methods are operating correctly.}

We ran 2048 shots of this problem on both the simulator and the emulator ten times over, and then we computed the average ratio between the probabilities of $|11\rangle$ and $|01\rangle$ to estimate the value of $|x\rangle$.
We found that the simulator produced a result of
\begin{equation}
    \begin{pmatrix}
        0.0948283 \\ 0.9051717
    \end{pmatrix},
    \label{ex1sim}
\end{equation}
and the emulator produced a result of
\begin{equation}
    \begin{pmatrix}
        0.09411943 \\ 0.90588057
    \end{pmatrix}.
    \label{ex1em}
\end{equation}
Both results are very close to \eqref{ex1sol}, which gives a final witness that both our simulation method and \textit{especially} our emulation method created and sampled from the distribution that a perfect quantum computer would when running the HHL algorithm on this problem.

Finally, the most important result of this experiment is the run-time.
The simulator ran all 2048 shots on average in 2.20873 seconds.
This is approximately 0.00108 seconds per shot.
The emulator, on the other hand, ran all 2048 shots in 0.07670 seconds on average, which is a time of 0.00003 seconds per shot.
All experiments were run on a standard Linux laptop.
Here, it is already demonstrated that the method of emulation offers a notable time complexity reduction compared to the simulation method.

The second example problem we used to test our simulator and emulator was
\begin{equation}
    \begin{pmatrix}
        13/2 & -1/2 \\
        -1/2 & 13/2
    \end{pmatrix}\begin{pmatrix}
        x_1 \\ x_2
    \end{pmatrix}=\begin{pmatrix}
        0 \\ 1
    \end{pmatrix},
    \label{ex2}
\end{equation}
chosen because the eigenvalues of $A$ are $\lambda_1=6$ and $\lambda_2=7$, hence the ratio between the eigenvalues is $6:7$.
Three qubits are needed to express this relation ($m=3$), and an additional two ancilla qubits are required to decompose the Toffoli gates required by the $R(\lambda^{-1})$ step into one- and two-qubit gates.\footnote{This is a requirement imposed by the implementation of the Intel Quantum Simulator.}
Consequently, we performed a simulation using 7 qubits.

\begin{figure*}[t]
    \centering
    \begin{subfigure}{\figsize\textwidth}
        \centering
        \includegraphics[width=1\linewidth]{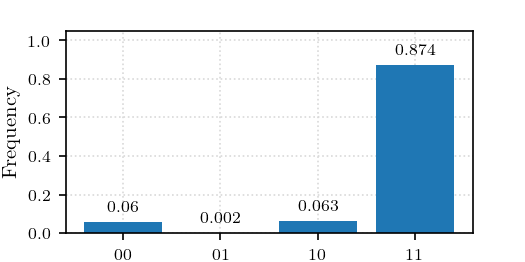}
        \caption{Simulation}
        \label{sim2}
    \end{subfigure}%
    \begin{subfigure}{\figsize\textwidth}
        \centering
        \includegraphics[width=1\linewidth]{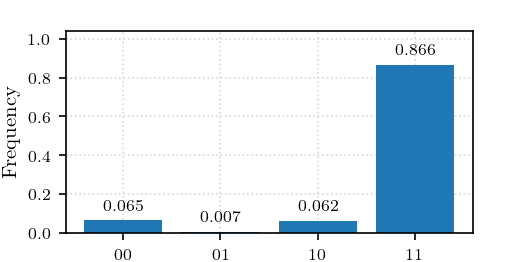}
        \caption{Emulation}
        \label{em2}
    \end{subfigure}
    \caption{Illustrative results of simulating and emulating the HHL algorithm on \eqref{ex2}.}
    \label{experiment2}
\end{figure*}

The solution to this equation is $x_1=1/84$ and $x_2=13/84$, which has a squared ratio of $0.00588235:0.99411765$ implying that $|x\rangle$ is
\begin{equation}
    \begin{pmatrix}
        0.00588235 \\ 0.99411765
    \end{pmatrix}.
    \label{ex2sol}
\end{equation}
Moreover, we ran 2048 shots of this problem on both the simulator and the emulator ten times over, and then estimated the value of $|x\rangle$.
We found that the simulator gave a solution of
\begin{equation}
    \begin{pmatrix}
        0.00452281 \\ 0.99547719
    \end{pmatrix},
    \label{ex2sim}
\end{equation}
and the emulator gave a solution of
\begin{equation}
    \begin{pmatrix}
        0.00657847 \\ 0.99342153
    \end{pmatrix}.
    \label{ex2em}
\end{equation}
The histogram for the simulator is given in Fig. \ref{sim2}, and the histogram for the emulator is given in Fig. \ref{em2}.

Once again, we find that the emulator greatly outperforms the simulator with respect to run-time.
The simulator ran all 2048 shots on average in 30.18495 seconds, which is approximately 0.01474 seconds per shot.
This demonstrates the exponential scaling of the simulator with respect to $m$---this required an order of magnitude more time than the $m=2$ problem.
The emulator, on the other hand, ran all 2048 shots in 0.07683 seconds on average, which is a time of 0.00003 seconds per shot.
This result is practically indistinguishable from the first experiment, demonstrating that the runtime of the emulator is invulnerable to changes in $m$.

These results experimentally validate that the runtime of the simulated HHL routine scales exponentially as a function of $n$ and $m$, while the emulated routine scales exponentially only as a function of $n$.
This represents an advancement in the state-of-the-art of executing the noiseless HHL algorithm.
 
\section{Error Analysis}

This work has an unusual relationship with the error inherent in the algorithm.
While most algorithms attempt to remove all errors from a given computation, we aim to mimic the behavior of a noiseless quantum computer running the HHL algorithm exactly.
This means that we attempt to replicate the error that would be observed on a real quantum device, while also eliminating any error that would not be observed on the same device.
Note that we are not concerned with simulating noise due to decoherence, control error, or other quantum processes; rather, we are concerned only with simulating the noise that is inherent to the ideal HHL algorithm.

Excluding hardware noise, there are two sources of error in the HHL algorithm.
The first source of error is when the ancilla qubit of the $a$ register is measured to be a $|0\rangle$ instead of a $|1\rangle$.
The algorithm \textit{minimizes} this error through the appropriate choice of the number of qubits in the $c$ register (which dictates the precision of the approximation of the ratio between the scaled eigenvalues of $A$), but this error source cannot be \textit{eliminated} in general \cite{harrow_quantum_2009}.
Therefore, because this error is observed on real devices, both our simulator and our emulator preserve this source of error.
This can be seen in Figs. \ref{experiment1} and \ref{experiment2}, where all plots show some probability mass on the states $|00\rangle$ and $|10\rangle$ (i.e., the states where $a=|0\rangle$).

The second source of error arises from the inaccuracy of the approximation of $|x\rangle$.
Recall that $|x\rangle$ is approximated by computing the ratio between the probability of measuring a $|11\rangle$ and the probability of measuring a $|01\rangle$.
As the number of shots for a given experiment increases, we expect that this error will tend toward zero (i.e., we expect that a perfect quantum computer will make an arbitrarily close approximation given enough iterations).
Thus, we hope to eliminate this source of error in our simulator and emulator.
To show that we have done this, we compute the absolute error between our approximations and the true values of $|x\rangle$ below.

For experiment 1, we have that the simulator solution, equation \eqref{ex1sim}, differs from the true solution, equation \eqref{ex1sol}, by a value of 0.01034.
The emulator solution \eqref{ex1em} differs from \eqref{ex1sol} by a value of 0.01709.
For experiment 2, the simulator solution \eqref{ex2sim} differs from the true solution \eqref{ex2sol} by a value of 0.00316, while the emulator solution \eqref{ex2em} differs from the true solution by a value of 0.00436.
The error is low on all accounts and is within the reasonable effects of shot noise.

As a final note of this section, let us consider the error between our two approximations.
Both the simulator and emulator aim to replicate the behavior of a real quantum device running the HHL algorithm, and hence both should produce the same output.
As a check of this, we compute the error between our two approximations.
We find that the mean absolute error between the two histograms of experiment 1 is 0.00500, and the mean absolute error between the two histograms of experiment 2 is 0.00475.
Again, both error rates are low enough to be explained by shot noise.
This adds a final witness that our two methods prepare and sample from the same target distribution, namely, the distribution prepared by an ideal quantum computer executing the HHL algorithm.

\section{Concluding Remarks}

In closing, our contribution to the field is an emulator for the HHL algorithm.
This represents an extension of the Unconventional Noiseless Intermediate Quantum Emulator (UNIQuE) \cite{Robertson_UNIQuE_2025}.
Our emulator efficiently replicates the behavior of a perfect quantum computer that executes the HHL algorithm on a linear system of equations $A|x\rangle=|b\rangle$.
An implementation for a $2\times2$ matrix $A$ is shown in the appendix.
Our emulator scales exponentially with the number of qubits required to represent $A$, which is an improvement over the standard state vector simulation method, which scales exponentially as a function of the number of qubits to both represent $A$ and approximate the ratio between the eigenvalues of $A$.

We validated the accuracy and speedup of our emulator through comparison with the Intel Quantum Simulator \cite{Guerreschi_2020}.
We performed two experiments using problems of varying complexity, and we verified that the emulator and the simulator agreed in each case.
Moreover, our first experiment reproduced the results of \cite{morrell2023stepbystep}, which added further validity to the accuracy of both our emulator and our simulator implementations.

\textbf{Data availability} All data and source code files for the project are available at \url{https://github.com/reecejrobertson/UNIQuE_HHL}.

\appendix

The following box gives a Python implementation of the emulation routine for a linear system with a $2\times2$ matrix $A$.






    

\begin{figure}[h]
    \centering
    \includegraphics[width=\linewidth]{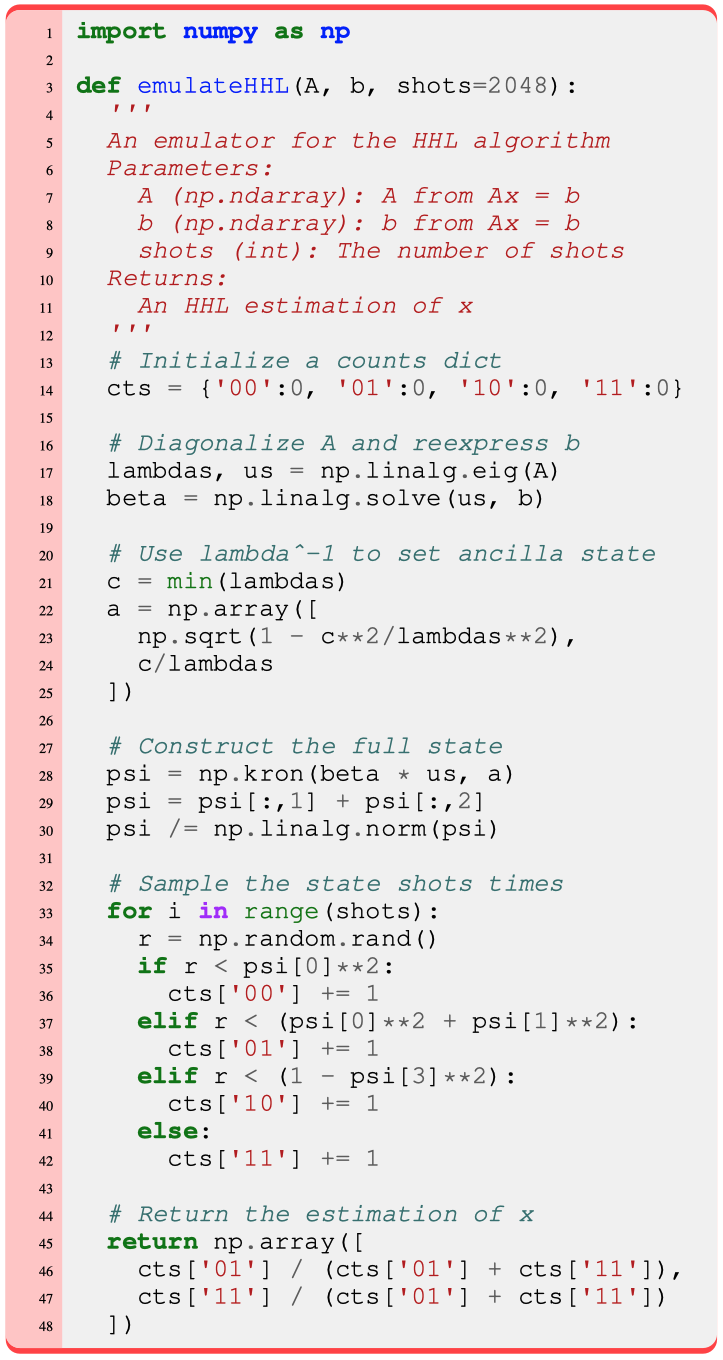}
\end{figure}

\printbibliography

\end{document}